\documentclass[twocolumn,showpacs,preprintnumbers,amsmath,amssymb]{revtex4}

\usepackage{graphicx}
\usepackage{dcolumn}
\usepackage{bm}

\begin{document}


\title{Spin accumulation created electrically in an $n$-type germanium channel using Schottky tunnel contacts}
\author{K. Kasahara,$^{1}$ Y. Baba,$^{1}$ K. Yamane,$^{1}$ Y. Ando,$^{1}$ S. Yamada,$^{1}$ Y. Hoshi,$^{2}$ K. Sawano,$^{2}$ M. Miyao,$^{1}$ and K. Hamaya$^{1,3}$\footnote{E-mail: hamaya@ed.kyushu-u.ac.jp} }
\affiliation{$^{1}$Department of Electronics, Kyushu University, 744 Motooka, Fukuoka 819-0395, Japan}
\affiliation{$^{2}$Advanced Research Laboratories, Tokyo City University, 8-15-1 Todoroki, Tokyo 158-0082, Japan}
\affiliation{$^{3}$PRESTO, Japan Science and Technology Agency, Sanbancho, Tokyo 102-0075, Japan}

\date{\today}

\begin{abstract}
Using high-quality Fe$_{3}$Si/$n^{+}$-Ge Schottky-tunnel-barrier contacts, we study spin accumulation in an $n$-type germanium ($n$-Ge) channel. In the three- or two-terminal voltage measurements with low bias current conditions at 50 K, Hanle-effect signals are clearly detected only at a forward-biased contact. These are reliable evidence for electrical detection of the spin accumulation created in the $n$-Ge channel. The estimated spin lifetime in $n$-Ge at 50 K is one order of magnitude shorter than those in $n$-Si reported recently. The magnitude of the spin signals cannot be explained by the commonly used spin diffusion model. We discuss a possible origin of the difference between experimental data and theoretical values.
\end{abstract}

\maketitle
For overcoming the scaling limit of silicon (Si) complementary metal oxide semiconductor (CMOS) devices,\cite{Antoniadis} germanium (Ge) channels with high electron and hole mobility have been expected.\cite{Tezuka,Lee,Miyao} In general, there are two critical issues for source-drain technologies of Ge-metal-oxide-semiconductor field effect transistors (MOSFETs), i.e., the strong Fermi-level pinning (FLP) at the metal/Ge interface\cite{Dimoulas,Toriumi} and low solubility of the dopants.\cite{Chui,Ikeda} Some solutions for the two issues have so far been proposed.\cite{Nishimura,Zhou,Thathachary,Yamane,Sawano} Recently, we individually addressed them by fabricating an atomically controlled metal/Ge interface\cite{Yamane} and forming an ultrashallow contact with the Sb $\delta$-doping.\cite{Sawano} 

If the Ge technologies for the high performance MOSFETs are combined with spintronic ones for the nonvolatile memory, one can realize a next-generation CMOS technology with ultra-low-power consumption.\cite{Wolf,Zutic} To date, spin-polarized electrons created by an optical method were detected electrically in Ge-based heterostructures.\cite{Shen} Very recently, electrical spin injection and detection in a Ge channel were also demonstrated in lateral devices with ferromagnet (Fe)-insulator (MgO or Al$_\text{2}$O$_\text{3}$) contacts.\cite{Zhou1,Saito,Jain,Jeon} Since the insulating barriers for source and drain contacts can result in large parasitic resistance unfortunately, a technique without the insulating barrier for the spin injection and detection in Ge will be required in the scaled MOSFET structures. 

In this paper, we show reliable evidence for the detection of spin accumulation created electrically in an $n$-Ge channel through Fe$_{3}$Si/$n^{+}$-Ge Schottky -tunnel-barrier contacts. This is the first step of research and development of Ge-based spintronic devices without using insulating tunnel contacts. 
\begin{figure}
\includegraphics[width=8.5cm]{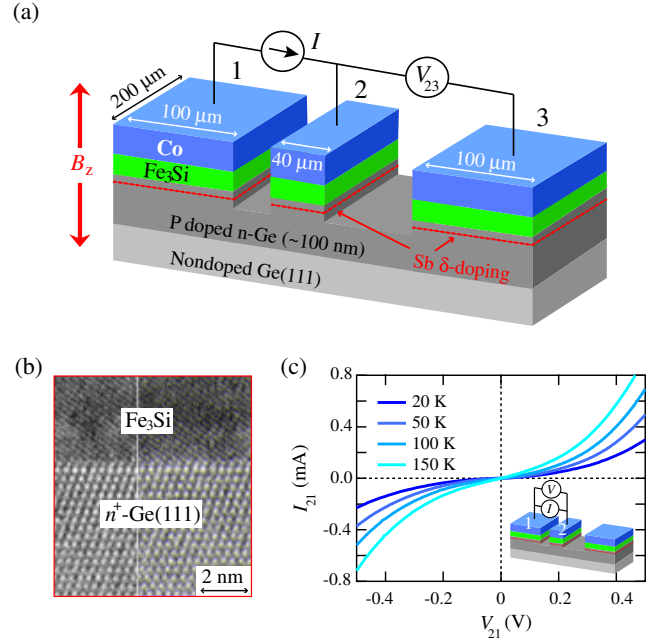}
\caption{(Color online) (a) Schematic diagram of a lateral three-terminal device with high-quality Fe$_{3}$Si/Ge contacts. (b) High-resolution transmission electron micrograph of the Fe$_{3}$Si/$n^{+}$-Ge interface. (c) $I-V$ characteristics measured between contacts 1 and 2 for various temperatures.} 
\end{figure}

Fe$_{3}$Si epitaxial films with a thickness of 10 nm were grown on Ge(111) by low-temperature molecular beam epitaxy (MBE) at 130 $^{\circ}$C.\cite{Hamaya} Prior to the growth, we fabricated a phosphorus-doped $n$-Ge(111) channel ($\sim$10$^{18}$ cm$^{-3}$) with a thickness of $\sim$100 nm on non-doped Ge(111) substrates ($\rho =$ $\sim$ 40 $\Omega$cm) by using an ion implantation technique and an annealing at 600 $^{\circ}$C.\cite{nGe} After the fabrication of the $n$-Ge(111) channel, the $n^{+}$-Ge(111) layer consisting of a 5-nm-thick Ge epilayer and $\delta$-doped Sb ($n$ = $\sim$10$^{14}$ cm$^{-2}$) was grown by MBE.\cite{Sawano} Conventional processes with photolithography, Ar$^{+}$ ion milling, and reactive ion etching were used to fabricate a three-terminal lateral device for measurements of the voltage changes induced by a Hanle effect.\cite{Lou,Tran,Dash,Sasaki,Ando,Dash2} Here a polycrystalline Co layer with a thickness of 25 nm was deposited on the Fe$_{3}$Si layer by using electron beam evaporation so as to align the magnetic moments in the in-plane direction.

The fabricated device is illustrated schematically in Fig. 1(a). As shown in the cross-sectional transmission electron micrograph of Fig. 1(b), the heterointerface consisting of Fe$_{3}$Si/$n^{+}$-Ge was atomically flat, leading to the reduction in the presence of interface states.\cite{Yamane,Sawano} Each contact denoted by 1, 2, or 3 has a lateral dimension of 100 $\times$ 200 $\mu$m$^{2}$, 40 $\times$ 200 $\mu$m$^{2}$ or 100 $\times$ 200 $\mu$m$^{2}$, respectively. The edge-edge distance between contacts 1 and 2 or contacts 2 and 3 was 10 $\mu$m or 25 $\mu$m, respectively, and the $n^{+}$-Ge(111) layer on the channel region was removed by Ar$^{+}$ ion milling. The Hanle-effect measurements were performed by a conventional dc method at 50 K. External magnetic fields ($B_\text{Z}$) for the Hanle-effect measurements were applied perpendicular to the film plane after the magnetic moments of all the contacts were aligned parallel to the film plane.
\begin{figure}
\includegraphics[width=8.5cm]{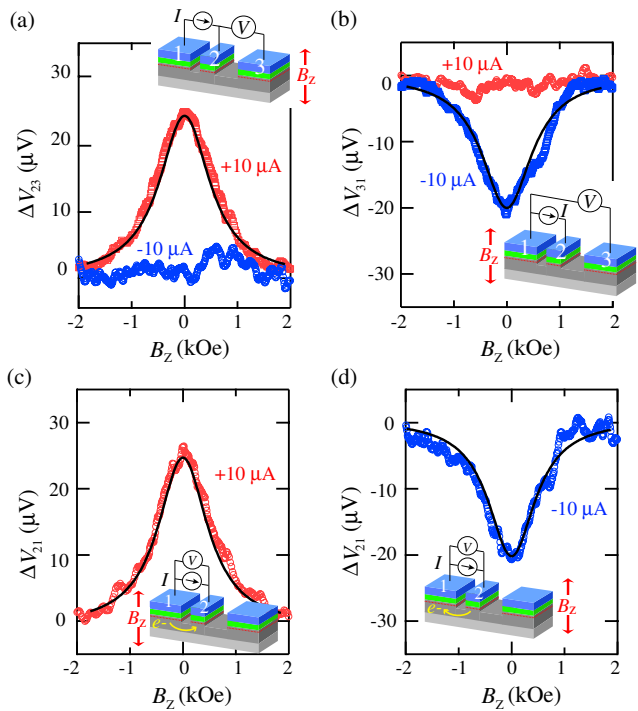}
\caption{(Color online) The three-terminal (a) $\Delta V_\text{23}$ and (b) $\Delta V_\text{31}$ versus $B_\text{Z}$ with current flows ($I_\text{21}$) of $\pm$10 $\mu$A at 50 K, measured with the terminal configurations schematically shown in the insets. Here $I_\text{21} =$ $+$10 $\mu$A indicates that electrons are injected from contact 1 into Ge and are extracted from Ge into contact 2. The two-terminal $\Delta V_\text{21}$ versus $B_\text{Z}$ with (c) $I_\text{21} =$ $+$10 $\mu$A and (d) $-$10 $\mu$A at 50 K.} 
\end{figure} 

Figure 1(c) shows two-terminal current-voltage ($I - V$) characteristics measured between contacts 1 and 2 for various temperatures. When the sign of $I$$_\text{21}$ is positive, the electrons are injected from the contact 1 into the Ge channel and are extracted from the Ge channel into the contact 2. We can see nonlinear curves for all the temperatures and a weak temperature dependence, indicating that tunneling conduction of electrons through the high-quality Fe$_{3}$Si/$n^{+}$-Ge/$n$-Ge interfaces is dominant. Thus, the asymmetry of the $I - V$ characteristics with respect to the $V_\text{21}$ polarity is quite small. 

Using this device, we measure the three-terminal voltage, $V$$_\text{23}$, as a function of $B_\text{Z}$ at 50 K in the terminal configuration shown in the inset schematic of Fig. 2(a). The red and blue plots show the data for $I$$_\text{21} =$ $+$10 and $-$10 $\mu$A, respectively, where a quadratic background voltage depending on $B_\text{Z}$ is subtracted from the raw data. For $I$$_\text{21} =$ $+$10 $\mu$A in Fig. 2(a) (red), a voltage change ($\Delta$$V_\text{23}$) of $\sim +$25 $\mu$V can be seen with increasing $B_\text{Z}$ from zero to $\pm$2 kOe. This is a consequence of the depolarization of spin-polarized electrons, i.e., Hanle-effect curve,\cite{Dash,Lou,Tran,Sasaki,Ando} indicating the first experimental detection of spin accumulation created electrically in $n$-Ge using Schottky tunnel contacts. In contrast, for $I$$_\text{21} =$ $-$10 $\mu$A (blue), we cannot see such voltage changes. This feature means that the presence of the spin accumulation is verified only by a forward-biased contact, i.e., contact 2 in the terminal configuration of Fig. 2(a). When we use the terminal configuration shown in the inset of Fig. 2(b), a voltage change ($\Delta$$V_\text{31}$) of $\sim -$20 $\mu$V can only be seen for $I$$_\text{21} =$ $-$10 $\mu$A in Fig. 2(b). This is also evidence for the presence of the spin accumulation in $n$-Ge, detected only by a forward-biased contact, i.e., contact 1. The asymmetric bias dependence can be understood by a difference in the spin-detection sensitivity through the Schottky-tunnel barrier for the spin accumulation in the conduction band of semiconductors, which has already been described by Lou {\it et al.} in $n$-GaAs channels.\cite{Lou} Recently, we also observed the same features in $n$-Si channels.\cite{Ando} 

In addition to the three-terminal methods, Lou {\it et al.} confirmed the above bias-dependent detectability by measuring two-terminal Hanle-effect curves in both bias polarities for GaAs-based devices.\cite{Lou} Following their experiments, we also confirm the two-terminal voltage, $V$$_\text{21}$, as a function of $B_\text{Z}$ at 50 K in Figs. 2(c) and 2(d). For $I$$_\text{21} =$ $+$10 $\mu$A or $-$10 $\mu$A, a voltage change ($\Delta$$V_\text{21}$) of $\sim +$25 $\mu$V or $\sim -$20 $\mu$V is observed, respectively. The magnitude of $V$$_\text{21}$, observed in Figs. 2(c) and 2(d), is almost equal to that shown in Figs. 2(a) and 2(b), respectively. Accordingly, the spin accumulation signals are detected only at the forward-biased contact in such low bias-current region. Considering these facts, we can judge that the spin accumulation created electrically in an $n$-Ge channel is evidently detected using Fe$_{3}$Si/Ge Schottky tunnel contacts. Here we also tried to measure the inverted Hanle signals, recently reported by Dash {\it et al}.\cite{Dash2} As a result, we could not see inverted Hanle signals at 50 K. This fact indicates that there is almost no local fluctuation of the magnetic fields at the atomically flat Fe$_{3}$Si/$n^{+}$-Ge/$n$-Ge interface.

A lower limit of spin lifetime ($\tau_\text{S}$) can be estimated from the Hanle-effect signals with a Lorentzian function,\cite{Dash} $\Delta$$V$($B_\text{Z}$) $=$ $\Delta$$V$(0)/[1+($\omega_\text{L}$$\tau_\text{S}$)$^{2}$], where $\omega_\text{L} =$ $g\mu_\text{B}$$B_\text{Z}$/$\hbar$ is the Lamor frequency, $g$ is the electron $g$-factor ($g =$ 1.563),\cite{Vrijen} $\mu_\text{B}$ is the Bohr magneton. The fitting curves (black solid curves) are shown in Figs. 2 (a), (b), (c), and (d), and the $\tau_\text{S}$ values are roughly estimated to be 125 $\sim$ 140 psec at 50 K, which are one order of magnitude shorter than those in Si channels at low temperatures reported recently.\cite{Sasaki,Ando} We can infer that the relatively short $\tau_\text{S}$ is arising from the presence of the strong spin-orbit interaction in Ge compared to Si.\cite{Hasegawa_Roth} 

Finally, we discuss the magnitude of $\Delta$$V$ (spin signals). For the spin signals with both contacts 1 ($A =$100 $\times$ 200 $\mu$m$^{2}$) and 2 ($A =$ 40 $\times$ 200 $\mu$m$^{2}$), we can simply estimate the spin resistance-area-product (spin-$RA$) at 50 K, $\frac{\Delta V}{I_\text{21}}$$\times$$A$, to be $\sim$ 4.0 $\times$ 10$^{4}$ $\Omega$$\mu$m$^{2}$ and $\sim$ 2.0 $\times$ 10$^{4}$ $\Omega$$\mu$m$^{2}$, respectively. Using the common diffusion model,\cite{Fert} we can also calculate the predicted spin signal as $\rho_\text{Ge}$($\lambda_{\rm Ge}$)$^{2}$/$w$ $\sim$ 61.9 $\Omega$$\mu$m$^{2}$, where $\rho_\text{Ge}$ (1.56 $\times$ 10$^{-3}$ $\Omega$cm) and $\lambda_{\rm Ge}$ ($\sim$ 0.63 $\mu$m)\cite{cal} are resistivity and spin diffusion length of Ge at 50 K, respectively, and $w$ (0.10 $\mu$m) is the thickness of the channel in the fabricated device. As a result, there are large differences (three orders of magnitude) between experimental data and theoretical values. In this context, we discuss two origins as follows. First, we can consider that the actual area associated with the tunneling of electrons is deviated largely from the fabricated contact area, arising from the in-plane inhomogeneous doping density near the interface between Fe$_{3}$Si and Ge. Actually, the distribution of the heavily doped Sb can become inhomogeneous under the Sb $\delta$-doping process.\cite{Sawano} Since the actual area associated to the tunneling conduction was quite small, the difference between the experiment and theory may be quite large. Second, two-step tunneling process via the localized states between Fe$_{3}$Si and Ge, as proposed by Tran {\it et al.},\cite{Tran} should be considered. However, we consider that the asymmetric bias dependence shown in Figs. 2 can be explained only by the difference in the spin-detection sensitivity for the spin accumulation in the Ge conduction band, as discussed in detail for Si-based devices.\cite{Ando} Therefore, we infer that the former is the one of the possible origins of the large difference between experimental data and theoretical values. We should further polish a technique of the Sb $\delta$-doping processes for fabricating the Schottky-tunnel-barrier contacts. 

In summary, we have obtained reliable evidence for the detection of spin accumulation created electrically in an $n$-Ge channel using high-quality Fe$_{3}$Si/$n^{+}$-Ge/$n$-Ge Schottky tunnel contacts in lateral structures. The estimated spin lifetime in $n$-Ge at 50 K was one order of magnitude shorter than those in $n$-Si reported recently. However, the magnitude of the spin signals was deviated largely from the theoretical value based on the spin diffusion model. One of the possible origins of the large deviation is the difference between the actual area associated with the tunneling conduction and the contact area of the fabricated device.

\vspace{2mm}
K.H. and M.M. acknowledge Prof. Y. Shiraki, Prof. H. Nakashima, and Prof. T. Kimura for useful discussions. Three of the authors (K.K.,Y.A., and S.Y.) acknowledge JSPS Research Fellowships for Young Scientists. This work was partly supported by Industrial Technology Research Grant Program from NEDO and Grant-in-Aid for Young Scientists (A) from JSPS. 

\noindent{{\bf References}}


\begin{thebibliography}{3}
\bibitem{Antoniadis}
D. A. Antoniadis, Proc. Symp. VLSI Technology, 2, (2002).
\bibitem{Tezuka}
T. Tezuka, S. Nakaharai, Y. Moriyama, N. Sugiyama, and S. Takagi, IEEE Electron Device Lett. {\bf 26}, 243 (2005).
\bibitem{Miyao}
M. Miyao, K. Toko, T. Tanaka, and T. Sadoh, Appl. Phys. Lett. {\bf 95}, 022115 (2009).
\bibitem{Lee}
C. H. Lee, T. Nishimura, T. Tabata, S. K. Wang, K. Nagashio, K. Kita, and A. Toriumi, IEDM Tech. Dig., 416 (2010).
\bibitem{Dimoulas}
A. Dimoulas, P. Tsipas, A. Sotiropoulos, and E. K. Evangelou, Appl. Phys. Lett. {\bf 89}, 252110 (2006).
\bibitem{Toriumi}
T. Nishimura, K. Kita, and A. Toriumi, Appl. Phys. Lett. {\bf 91}, 123123 (2007).
\bibitem{Chui}
C. O. Chui, K. Gopalakrishnan, P. B. Griffin, J. D. Plummer, and K. C. Saraswat, Appl. Phys. Lett. {\bf 83}, 3275 (2003).
\bibitem{Ikeda}
K. Ikeda, Y. Tamashita, N. Sugiyama, N. Taoka, and S. Takagi, Appl. Phys. Lett. {\bf 88}, 152115 (2006).
\bibitem{Nishimura}
T. Nishimura, K. Kita, and A. Toriumi, Appl. Phys. Exp. {\bf 1}, 051406 (2008).
\bibitem{Zhou}
Y. Zhou, W. Han, Y. Wang, F. Xiu, J. Zou, R. K. Kawakami, and K. L. Wang, Appl. Phys. Lett. {\bf 96}, 102103 (2010).
\bibitem{Thathachary}
A. V. Thathachary, K. N. Bhat, N. Bhat, and M. S. Hegde, Appl. Phys. Lett. {\bf 96}, 152108 (2010).
\bibitem{Yamane}
K. Yamane, K. Hamaya, Y. Ando, Y. Enomoto, K. Yamamoto, T. Sadoh, and M. Miyao, Appl. Phys. Lett. {\bf 96}, 162104 (2010).
\bibitem{Sawano}
K. Sawano, Y. Hoshi, K. Kasahara, K. Yamane, K. Hamaya, M. Miyao, and Y. Shiraki, Appl. Phys. Lett. {\bf 97}, 162108 (2010).
\bibitem{Wolf}
S. A. Wolf, D. D. Awschalom, R. A. Buhrman, J. M. Daughton, S. v. Moln\'ar, M. L. Roukes, A. Y. Chtchelkanova, and D. M. Treger, Science {\bf 294}, 1488 (2001).
\bibitem{Zutic}
I. \v{Z}utic, J. Fabian, and S. D. Sarma, Rev. Mod. Phys. {\bf 76}, 323  (2004).
\bibitem{Shen}
C. Shen, T. Trypiniotis, K. Y. Lee, S. N. Holmes, R. Mansell, M. Husain, V. Shah, X. V. Li, H. Kurebayashi, I. Farrer, C. H. de Groot, D. R. Leadley, G. Bell, E. H. C. Parker, T. Whall, D. A. Ritchie, and C. H. W. Barnes, Appl. Phys. Lett. {\bf 97}, 162104 (2010).
\bibitem{Zhou1}
Y. Zhou, W. Han, L.-T. Chang, F. Xiu, M. Wang, M. Oehme, I. A. Fischer, J. Schulze, R. K. Kawakami, and K. L. Wang, Phys. Rev. B {\bf 84}, 125323 (2011). 
\bibitem{Saito}
H. Saito, S. Watanabe, Y. Minenob, S. Sharma, R. Jansen, S. Yuasa, K. Ando, Solid State. Commun {\bf 151}, 1159 (2011).
\bibitem{Jain}
A. Jain, L. Louahadj, J. Peiro, J. C. Le Breton, C. Vergnaud, A. Barski, C. Beign\'e, L. Notin, A. Marty, V. Baltz, S. Auffret, E. Augendre, H. Jaffr\`es, J. M. George, M. Jamet, arXiv:1107.3510v1.
\bibitem{Jeon}
K. Jeon, B. Min, Y. Jo, H. Lee, I. Shin, C. Park, S. Park and S. Shin, arXiv:1108.3145v1.
\bibitem{Hamaya}
T. Sadoh, M. Kumano, R. Kizuka, K. Ueda, A. Kenjo, and M. Miyao, Appl. Phys. Lett. {\bf 89}, 182511 (2006); K. Hamaya, Y. Ando, T. Sadoh, and M. Miyao, Jpn. J. Appl. Phys. {\bf 50}, 010101 (2011); K. Hamaya, T. Murakami, S. Yamada, K. Mibu, and M. Miyao, Phys. Rev. B {\bf 83}, 144411 (2011).
\bibitem{nGe}
C. O. Chui, L. Kulig, J. Moran, W. Tsai, and K. Saraswat, Appl. Phys. Lett. {\bf 87}, 091909 (2005); A. Satta, T. Janssens, T. Clarysse, E. Simoen, M. Meuris, A. Benedetti, I. Hoflijk, B. De Jaeger, C. Demeurisse, and W. Vandervorstb, J. Vac. Sci. Technol. B {\bf 24}, 494 (2006). 
\bibitem{Lou}
X. Lou, C. Adelmann, M. Furis, S. A. Crooker, C. J. Palmstr\o m, and P. A. Crowell, Phys. Rev. Lett. {\bf 96}, 176603 (2006).
\bibitem{Tran}
M. Tran, H. Jaffr\`es, C. Deranlot, J. -M. George, A. Fert, A. Miard, and A. Lema\^{\i}tre, Phys. Rev. Lett. {\bf 102}, 036601 (2009).
\bibitem{Dash}
S. P. Dash, S. Sharma, R. S. Patel, M. P. Jong, and R. Jansen, Nature (London) {\bf 462}, 491 (2009); S. P. Dash, S. Sharma, J. C. Le Breton, and R. Jansen, Proc. SPIE 7760, 77600J (2010).
\bibitem{Sasaki}
T. Sasaki, T. Oikawa, M. Shiraishi, Y. Suzuki, and K. Noguchi, Appl. Phys. Lett. {\bf 98}, 012508 (2011).
\bibitem{Ando}
Y. Ando, K. Kasahara, K. Yamane, Y. Baba, Y. Maeda, Y. Hoshi, K. Sawano, M. Miyao, and K. Hamaya, Appl. Phys. Lett. {\bf 99}, 012113 (2011).
\bibitem{Dash2}
S. P. Dash, S. Sharma, J. C. Le Breton, H. Jaffr\`es, J. Peiro, J.-M. George, A. Lemaitre, and R. Jansen, Phys. Rev. B {\bf 84}, 054410 (2011).
\bibitem{Vrijen}
R. Vrijen, E. Yablonovitch, K. Wang, H. W. Jiang, A. Balandin, V. Roychowdhury, T. Mor, and, D. DiVincenzo, Phys. Rev. A {\bf 62}, 012306 (2000).
\bibitem{Hasegawa_Roth}
H. Hasegawa, Phys. Rev. {\bf 118}, 1523 (1960); L. M. Roth, Phys. Rev. {\bf 118}, 1534 (1960).
\bibitem{Fert}
A. Fert and H. Jaffr\`es, Phys. Rev. B {\bf 64}, 184420 (2001); A. Fert, J.-M. George, H. Jaffr\`es, and R. Mattana, IEEE Trans. Electron Devices {\bf 54}, 921 (2007).
\bibitem{cal}
$\lambda_{\rm Ge}$ ($\sim$ 0.63 $\mu$m) is calculated by $\sqrt{D\tau_\text{S}}$, where $D$ is the diffusion constant of our channel at 50 K, $D \sim$ 27.7 cm$^{2}$s$^{-1}$. For our device, the width ($W$) of the contact 1 and 2  is quite lager than $\lambda_{\rm Ge}$.


\end{thebibliography}
\end{document}